\documentclass[11pt]{article}
\usepackage{url}
\usepackage{bbm}
\usepackage{amsthm}
\usepackage{amsmath}
\usepackage{subfigure}
\usepackage{bbm}
\usepackage{amsthm}
\usepackage{amsmath}
\usepackage{caption}
\usepackage{graphicx}
\usepackage{multirow}
\usepackage{xcolor}
\usepackage{natbib}
\usepackage[margin=1in]{geometry}
\usepackage{authblk}

\begin{document}

\title{Sequential Control of False Positives in Online Change Point Detection}

\author{MELISSA LYNNE MARTIN$^{1}$, THEODORE D. SATTERTHWAITE$^{2,3,4}$, IAN J. BARNETT$^1$\\
\textit{$^1$Department of Biostatistics, Epidemiology, and Informatics,
University of Pennsylvania,
Philadelphia, PA,
USA; $^2$Department of Psychiatry,
University of Pennsylvania,
Philadelphia, PA,
USA; $^3$Penn Lifespan Informatics and Neuroimaging Center (PennLINC),
University of Pennsylvania,
Philadelphia, PA,
USA; $^4$Penn-Children's Hospital of Philadelphia Lifespan Brain Institute,
University of Pennsylvania,
Philadelphia, PA,
USA}
\\
{martin30@pennmedicine.upenn.edu}}

\date{}


\maketitle


\begin{abstract}
{Online change point detection is the process of identifying distributional changes in time-ordered data in real time. In applications such as mobile health (mHealth), repeated testing is often performed as new data arrive, creating a multiple testing problem. Traditional approaches for controlling the family-wise error rate (FWER) are not well suited to this setting because the tests are highly dependent and the number of tests is not fixed in advance. In this work, we introduce a sequential family-wise error rate (sFWER), defined as the probability of at least one false positive within a moving monitoring window. We propose a simulation-based calibration procedure to estimate monitoring thresholds that control the sFWER at a desired level. Through simulation studies, we demonstrate that the proposed procedure achieves the desired error control, while commonly used alternatives are either overly conservative or fail to adequately control false alarms. Finally, we illustrate the proposed approach using passively collected smartphone data from a cohort of adolescents and young adults with affective instability.}
\end{abstract}

\section{Introduction}

Real-time monitoring involves assessing data streams where the data arrive sequentially over time. This is especially useful in applications where changes in data streams need to be identified in real-time soon after the change onset. One application area where this is used is in mobile health (mHealth), which uses smartphones and wearable devices to collect behavioral or health data over time. In psychiatric populations, mHealth can be used as a supplemental tool when outpatient psychotherapy and psychiatric medication management are insufficient for mitigating the risk of an adverse event (\citealp{torous_realizing_2015,torous_new_2016,mohr_personal_2017,onnela_harnessing_2016}). In these settings, a key goal is to detect changes in a patient's behavior as early as possible so a meaningful intervention can be deployed if necessary, motivating the use of online change point detection.

Online change point detection is the process of finding distributional changes in time-ordered data in real-time (\citealp{aminikhanghahi_survey_2017}). In practice, this requires performing sequential hypothesis tests over time as data accumulates. This creates a multiple testing problem, since we repeatedly test for a change point as new data arrive. Repeated testing over time creates a practical challenge in online change point detection. In many applications, investigators must balance sensitivity to true changes against the risk of false alarms. This is especially important in mHealth settings, where false alarms may lead to unnecessary follow-up or contribute to alert fatigue among individuals responsible for monitoring incoming alerts (\citealp{hravnak_call_2018,chaparro_reducing_2020}). Therefore, it is important to understand and control the probability of falsely detecting a change point during a clinically meaningful monitoring period.

A common goal in multiple testing is to control the probability of at least one false alarm, where a false alarm corresponds to a Type I error. In the change point detection setting, this means detecting a change point when there truly is no change point. This probability is commonly referred to as the family-wise error rate (FWER). However, the online change point detection setting differs from standard multiple testing problems because the tests are highly dependent. Specifically, the test statistics are computed using overlapping windows of data, which yields strong correlation of test statistics over time and makes error control more challenging. Another challenge in sequential testing is that the number of tests is not fixed. FWER is typically controlled over a set number of tests, but in our setting we continue testing until a change is detected. Therefore, traditional multiple testing corrections are not directly applicable, motivating the need for alternative approaches to control false positives in online monitoring settings.

There are many existing methods for controlling FWER in multiple testing, though they are not directly applicable for control of sequential, repeated testing settings. The Bonferroni correction is a widely used approach for controlling the FWER by adjusting significance thresholds according to the number of tests performed (\citealp{bonferroni_teoria_1936}), but it can be overly conservative when the tests are correlated. The Šidák correction (\citealp{sidak_rectangular_1967}) is slightly less conservative, but is still too conservative to use when testing highly correlated hypothesis tests in a sequential testing setting.

Several approaches have been proposed for controlling Type I error in sequential testing settings. In the clinical trials literature, group sequential methods allow interim analyses to be conducted while maintaining control of the overall Type I error rate (\citealp{pocock_group_1977,obrien_multiple_1979}). Alpha-spending approaches further extend these ideas by allowing greater flexibility in the timing of interim analyses while preserving error control (\citealp{lan_discrete_1983}). However, these methods are typically designed for settings with a finite number of planned analyses and a prespecified monitoring schedule. In contrast, online change point detection involves continual monitoring and repeated testing until a change is detected. These characteristics make existing sequential testing procedures difficult to apply directly and motivate the development of methods tailored to online monitoring settings.

In this work, we introduce a sequential family-wise error rate (sFWER), defined as the probability of at least one false alarm within a moving monitoring window. We then develop a simulation-based procedure for calibrating monitoring thresholds to control this error rate in online change point detection. Through simulation studies, we show that the proposed method achieves the desired error control, while standard approaches are either overly conservative or fail to adequately control false alarms. Finally, we apply the procedure to smartphone-based mHealth data to illustrate its use in practice.

\section{Methods} \label{sec:methods}

\subsection{Sequential testing problem} \label{sec:seq_testing_prob}

Consider an online change point detection setting where we let $r$ be the number of run-in days before daily sequential testing begins. On each day $t>r$, test statistic $S_t$ is computed and the null hypothesis $H_{0t}$, corresponding to no change point detected on day $t$, is tested. Consider a moving time window of length $\Delta$ over which a practitioner wants to control the probability of a false alarm, where a false alarm corresponds to rejecting $H_{0t}$. For example, a practitioner might want to control the probability of having at least one false alarm over a $\Delta=30$-day period. 

In traditional multiple testing settings, the family-wise error rate (FWER) is defined as the probability of at least one false alarm over a fixed set of tests. In the change point detection context, controlling the FWER means controlling the probability of at least one false positive across all monitoring days. However, in online change point detection, testing is sequential and ongoing until a change is detected, so the number of tests is itself random. To quantify false alarm rates in this setting, we define a sequential family-wise error rate (sFWER) as the probability of at least one false positive within a moving monitoring window. This differs from the classical FWER, which is defined over a fixed number of tests, by focusing on false positives occurring within a clinically meaningful monitoring period.

We assume that smaller values of test statistic $S_t$ provide stronger evidence against the global null hypothesis of no change point. Then to control sFWER at level $\alpha_\Delta$, we seek a cutoff $c$ such that for any $t \geq r$,

\begin{equation}
    Pr\left(\bigcup_{j=t+1}^{t+\Delta} \{S_{j} \leq c\}\right) \leq \alpha_\Delta
\end{equation}

Note that we require this bound to hold for any window of length $\Delta$ after the run-in period, rather than only the first $\Delta$ monitor days. This definition ensures that, at any point during monitoring, the probability of at least one false alarm in the next $\Delta$ days is controlled. This inequality is equivalent to

\begin{equation}
        Pr\left(\min_{t+1 \leq j \leq t+\Delta} S_j > c\right) \geq 1 - \alpha_\Delta
\end{equation}

The primary challenge in this setting is that the joint distribution of the sequential test statistics $S_t$, and therefore the distribution of $\min_{t+1 \leq j \leq t+\Delta} S_j$, is non-trivial. This is because the statistics are highly temporally dependent due to the overlapping data used across monitoring days. As a result, deriving an analytic cutoff $c$ that satisfies the above condition is generally intractable.

\subsection{Proposed calibration procedure} \label{sec:prop_calibr_proc}

We propose a simulation-based method for calibrating the test statistic cutoff $c$. Our approach is similar in spirit to resampling-based max-statistic methods for FWER control (e.g., \citealp{westfall_adjusting_1993}), but it is tailored to our online monitoring setting, where test statistics are computed sequentially and are inherently dependent.

Our goal is to estimate a cutoff $c$ such that the probability of at least one false alarm within a monitoring window of length $\Delta$ is controlled at the level $\alpha_\Delta$. From the formulation in the previous section, this reduces to understanding the distribution of

\begin{equation}
    M = \min_{t+1 \leq j \leq t+\Delta} S_j
\end{equation}

which represents the minimum test statistic over this window. Since this distribution is not available in closed form, we approximate it using simulations under the global null hypothesis of no change point. Specifically, we generate $B$ datasets of length $r+\Delta$ under the null hypothesis. For each dataset, we compute the sequence of monitoring statistics $\{S_{r+1}, \dots, S_{r+\Delta}\}$, and then compute the minimum over this window. This yields an empirical distribution of the window minimum under the null. We define the cutoff $c$ as the empirical $\alpha_\Delta$ quantile of this distribution.

In practice, during monitoring, we reject $H_{0t}$ and declare a change point at day $t$ if $S_t \leq c$. By construction, this procedure controls the probability of at least one false alarm within any monitoring window of length $\Delta$ at approximately $\alpha_\Delta$.

\subsection{VC* test statistic for online change point detection} \label{sec:vcstar}

In this paper we use VC*, a variance component score test for online multivariate change point detection (\citealp{martin_variance_2026}). On each monitoring day $t$, we consider candidate change point locations $k \in \{t-\mathrm{db}, \dots, t-1\}$, where $\mathrm{db}$ is the number of days back over which we search for a change point on each monitor day. For each candidate day $k$, we compute test statistic $Q_k$ based on a variance component score test for a change in the mean and/or covariance structure of the data. We convert each $Q_k$ to a p-value $p_k$ using a permutation-based approach, where $p_k$ represents the probability of observing a test statistic at least as extreme as $Q_k$ under the null hypothesis. We then define the monitoring statistic at day $t$ to be

\begin{equation}
    S_t = \min_{t-\mathrm{db} \leq k \leq t-1} p_k
\end{equation}
This definition places the candidate statistics on a common scale across candidate days and monitoring days. Smaller values of $S_t$ indicate stronger evidence against the null hypothesis of no change point.

Since $S_t$ is defined as the minimum p-value across candidate change point days, we do not expect the selected candidate day to be uniformly distributed, even under the null hypothesis of no change point. Candidate change point days near the edges of the search window are less correlated with the remaining candidate days than those in the middle, which makes them more likely to attain the minimum p-value by chance alone. As a result, the selected candidate day exhibits a U-shaped distribution under the null. This behavior is driven by the dependence structure induced by the overlapping data used to compute the candidate test statistics. An illustration of this phenomenon is provided in Supplementary Section S1.

\section{Simulations} \label{sec:sims}

We conduct simulation studies to evaluate the operating characteristics of the proposed monitoring procedure. We first explore the effect of the number of run-in days before change point detection begins on the resulting calibrated cutoff. We then investigate control of the sFWER under the null and the ability to detect true change points under alternative scenarios.

\subsection{Effect of run-in length on the calibrated cutoff} \label{sec:runin}

We first assess whether the calibrated cutoff generated from the proposed procedure is sensitive to the length of the run-in period $r$. This is important as the run-in length determines how much baseline data is needed before online change point detection begins.

We generate data under the global null hypothesis of no change point. Specifically, for $t = 1, ..., T$ we generate $Y_t \sim N_p(0,I_p)$ where all observations are generated independently over time. We consider both univariate and multivariate settings with $p \in \{1,10\}$ features. We evaluate run-in lengths $r \in \{30, 60, 100\}$, monitoring window lengths $\Delta \in \{7,14\}$, and target sFWER levels $\alpha_\Delta \in \{0.2, 0.1\}$. For each setting, we generate $B=1000$ null datasets of length $T=r+\Delta$, where the first $r$ observations correspond to the run-in period and the remaining $\Delta$ observations correspond to the monitoring window. We fix $\mathrm{db}=7$ in all settings.

At each monitoring day, we compute the VC* test statistic described in Section \ref{sec:vcstar} using permutation-based p-values across the previous $\mathrm{db}=7$ candidate change point days. For each simulated data set $b\in\{1,\dots, B\}$, we compute the sequence of test statistics $S_{r+1}^{(b)},...,S_{r+\Delta}^{(b)}$ and record the minimum test statistic over the monitoring window,

\[
M^{(b)} = \min_{r + 1 \le t \le r + \Delta} S_t^{(b)}
\]
The calibrated cutoff $c$ is defined as the empirical $\alpha_\Delta$ quantile of the simulated distribution of $\{M^{(1)},\dots, M^{(B)}\}$. We additionally use bootstrap resampling to assess the variability of the estimated cutoff.

\begin{figure}[ht]
\centering
\includegraphics[width=\textwidth]{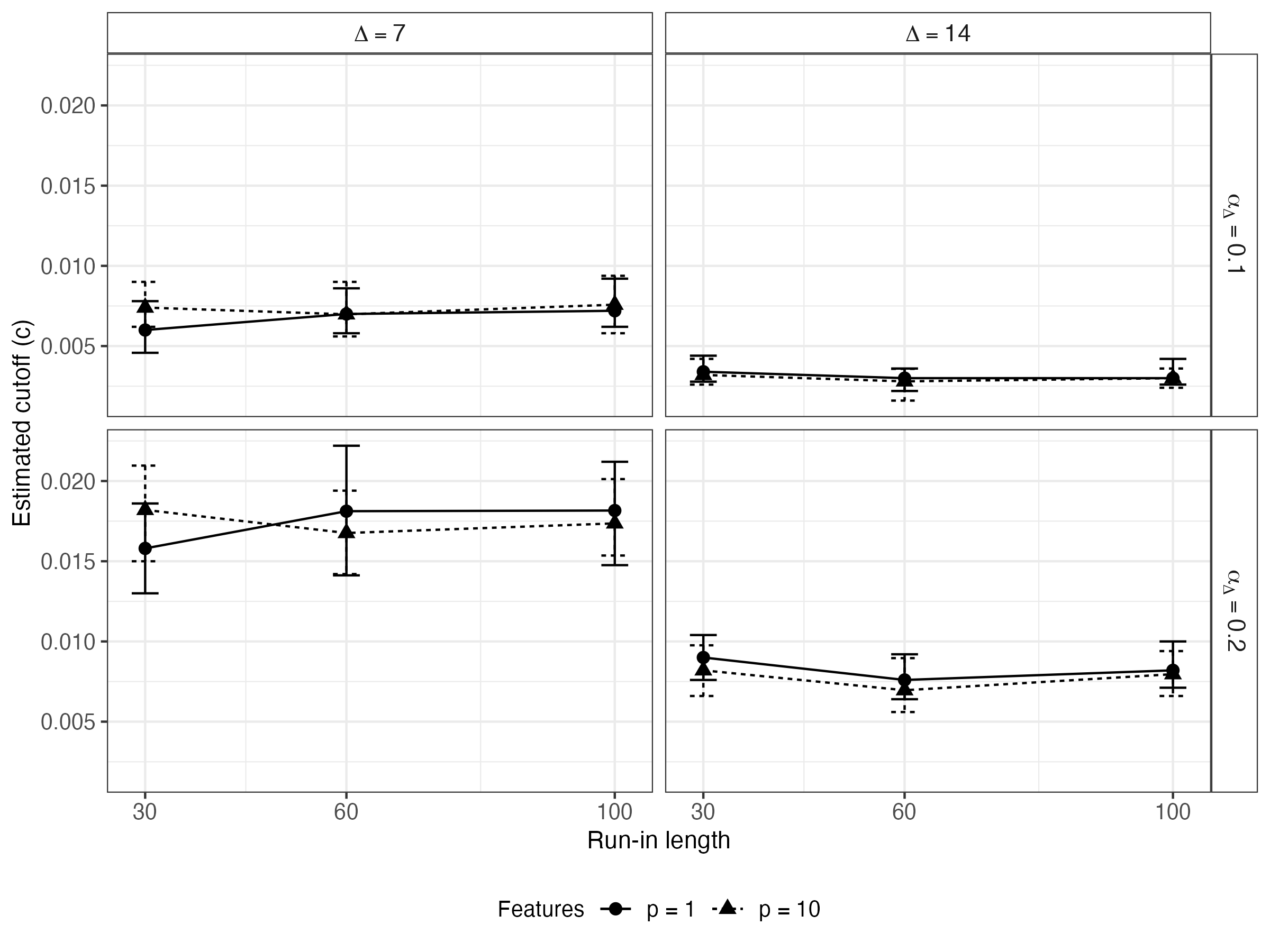}
\caption{Estimated calibrated cutoffs across run-in lengths. Points show the estimated cutoff, and error bars show the variability in the estimated cutoff across bootstrap samples. Separate panels correspond to combinations of monitoring window length $\Delta$ and target sFWER level $\alpha_\Delta$.}
\label{fig:runin_cutoffs}
\end{figure}

Figure \ref{fig:runin_cutoffs} shows the estimated cutoffs across run-in lengths for each setting. Overall, the estimated cutoffs are relatively stable across values of $r$, with substantial overlap in the bootstrap variability bands. The cutoffs are also similar between the univariate ($p=1$) and multivariate ($p=10$) settings. 

As expected, the calibrated cutoff depends on both $\Delta$ and $\alpha_\Delta$. For a fixed value of $\alpha_\Delta$, the cutoff is smaller when $\Delta=14$ than when $\Delta=7$. This reflects the fact that as $\Delta$ increases, there are more opportunities to observe an extreme minimum p-value over a longer monitoring window length. Similarly, for a fixed value of $\Delta$, the cutoff is smaller when $\alpha_\Delta=0.1$ than when $\alpha_\Delta=0.2$, which is consistent with a more stringent rejection threshold.

Since the proposed monitoring procedure rejects the null hypothesis when $S_t \leq c$, these results suggest that changing the run-in length would not substantially change the resulting sFWER or power. Moreover, since larger values of $r$ require additional baseline observations to be collected before monitoring can begin, we use the smaller run-in length of $r=30$ in the remaining simulation studies.

\subsection{sFWER and power simulations}

Next we conduct simulations to evaluate the sFWER and power of the proposed procedure and compare its performance with competing methods.

We consider monitoring windows of length $\Delta \in \{7,14\}$ and both univariate and multivariate settings with $p \in \{1,10\}$ features. We evaluate performance at target sFWER levels $\alpha_\Delta \in \{0.1,0.2\}$. Monitoring began after a run-in period of length $r=30$, with $\mathrm{db}=7$ candidate change point days evaluated at each monitoring time.

For each monitoring day $t>r$, we compute the statistic $S_t$ as described in Section \ref{sec:vcstar}. Specifically, for each candidate change point day in the previous $\mathrm{db} = 7$ days, we compute a permutation-based VC* p-value using 5000 resamples based only on data observed up to time $t$, and define $S_t$ as the minimum of these p-values.

For each combination of $\Delta$ and $\alpha_\Delta$, we used the calibrated cutoff obtained from the run-in length study in Section \ref{sec:runin}. Specifically, we used the cutoff corresponding to $r=30$, as the previous simulations suggested that the calibrated cutoff is not sensitive to the choice of run-in length. To evaluate sFWER, we generate an additional $B=1000$ null datasets of length $r+\Delta$ and record whether at least one false alarm occurred during the monitoring window, i.e., whether $M \leq c$. The estimated sFWER is computed as the proportion of simulated datasets with at least one rejection.

To assess detection performance under the alternative, we generate $B=1000$ datasets with a true mean shift of magnitude $\tau \in \{0.5,1.0,1.5\}$ occurring immediately after the run-in period (i.e., day $r+1$) and estimate the probability of detecting at least one change point during the monitoring window.

We compare the proposed procedure with three alternative approaches for controlling Type I error. The unadjusted procedure applies a fixed threshold of $\alpha=0.05$ at each monitoring day. Bonferroni and Šidák corrections are traditionally used to control the FWER across a fixed number of tests. To adapt these procedures to our sequential monitoring setting, we use $\Delta$, the length of the monitoring window, in place of the number of tests when computing the Bonferroni and Šidák thresholds. This results in per-day significance thresholds of $\alpha_\Delta/\Delta$ and $1-(1-\alpha_\Delta)^{1/\Delta}$, respectively.

Table \ref{tab:rejection_combined} summarizes rejection rates across all simulation settings. Under the null settings ($\tau=0$), the proposed procedure achieves rejection rates close to the target sFWER levels across both univariate and multivariate settings. Since the cutoff is estimated from a finite number of simulations, the observed sFWER is not expected to equal the target level exactly and may fall slightly above or below the target due to Monte Carlo variability. In contrast, the unadjusted procedure becomes increasingly liberal as the monitoring window increases, particularly for $\Delta=14$, where rejection rates substantially exceed the target level. Bonferroni and Šidák corrections, on the other hand, tend to be overly conservative, with rejection rates that often fall below the desired level.

Under the alternative ($\tau>0$), the proposed method maintains substantially higher detection rates than Bonferroni and Šidák across all settings, while still preserving the desired error control. As expected, detection rates increase with larger signal magnitudes (i.e., as $\tau$ increases). Compared with the unadjusted procedure, the proposed method achieves similar detection performance in many settings while avoiding the inflated false alarm rates seen under the null. The Bonferroni and Šidák corrections are conservative which leads to lower power when there is truly a change point. Overall, these results suggest that the proposed calibration approach provides a better balance between controlling false alarms and maintaining sensitivity to true changes than standard multiple testing corrections. These patterns are consistent across both univariate ($p=1$) and multivariate ($p=10$) settings.

\section{Real data application}
We apply the proposed monitoring procedure to passively collected smartphone data from a cohort of adolescents and young adults enrolled in a study investigating affective instability in youth (\citealp{xia_mobile_2022}). We follow the same preprocessing steps described in \citealp{martin_variance_2026}, summarized here.

\subsection{Participants and data acquisition}\label{sec:participants}
A total of 41 adolescents and young adults were enrolled in a study investigating affective instability in youth (\citealp{xia_mobile_2022}). Descriptive statistics summarizing participant diagnoses are provided in Xia et al. (2022).

Smartphone sensor and social data were acquired through the Beiwe platform \citep{torous_new_2016}. Once the Beiwe application was installed on participants' smartphones, Global Positioning System (GPS) and social data were passively collected and summarized into daily features. In this paper, we focus on continuous mobility-based features derived from GPS data. Computation of these daily summary features is described in the Supplement of \citealp{barnett_inferring_2020}.

\subsection{Data processing and analysis} \label{sec:data_proc_and_analysis}
Following the preprocessing steps described in \citealp{martin_variance_2026}, we consider segments of sensor data spanning at least 14 days with no more than three consecutive days of missingness. Participants could contribute more than one eligible segment, which were analyzed separately. Missing observations were imputed using the na.interp() function in R. Mobility features were then transformed and adjusted for day-of-week effects, and the resulting residuals were used as input to the change point detection procedure. Depending on data availability, segments contained either 11 or 13 mobility features.

For each data segment, we mimic an online implementation of change point detection using VC*. We begin with a run-in period of $r=14$ days, which is used to estimate the patient's baseline behavior. At each monitoring day $t > r$, we compute the VC* test statistic for each candidate change point day $k \in \{t-\mathrm{db}, \dots, t-1\}$ using only data observed up to time $t$. We set $\mathrm{db}=7$ and use a regularization parameter of 0.1 when estimating the covariance matrix. For each candidate day, we obtain a permutation-based p-value using 5000 resamples. We then define the monitoring statistic as
\[
S_t = \min_{t-\mathrm{db} \leq k \leq t-1} p_k,
\]
where $p_k$ is the p-value corresponding to candidate change point day $k$. 

To estimate a common test statistic cutoff across subjects, we first compute a correlation matrix for each subject using all available data. We then combine these subject-level correlation matrices into a sample-wide correlation matrix using a weighted average, where weights are proportional to the number of observed days contributed by each subject. Simulation results (see Supplement Section S2) suggest that the cutoff is relatively insensitive to the underlying correlation structure. Therefore, we use the pooled correlation matrix to generate data under the null and estimate a cutoff to control the sFWER at $\alpha_\Delta=0.2$ over a monitoring window of length $\Delta=7$. Because some segments contained 11 mobility features while others contained 13, cutoffs were calibrated separately for each feature dimension and applied accordingly during monitoring.

After the run-in period, monitoring proceeds sequentially. At each $t>r$, we test whether a change point has occurred within the previous $\mathrm{db}=7$ days. Monitoring continues until a change point is detected, at which point the procedure is reset and a new run-in period of $r=14$ days is initiated. We apply the proposed monitoring procedure using the calibrated cutoff corresponding to $\alpha_\Delta=0.2$ and $\Delta=7$.


\subsection{Results}
\begin{figure}[ht]
\centering
\includegraphics[width=\textwidth]{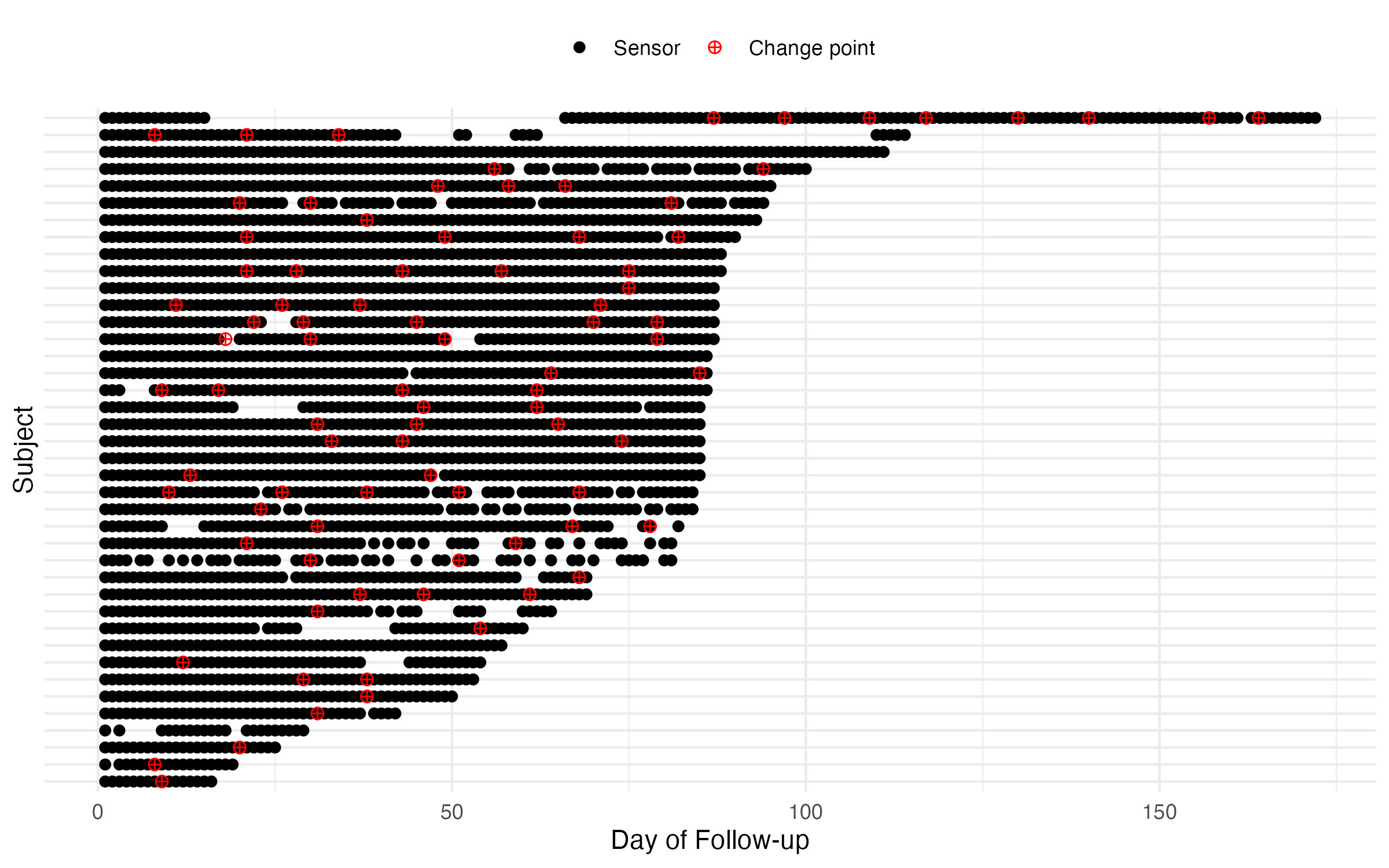}
\caption{Detected change points for each participant using the proposed monitoring procedure. Black points indicate days on which smartphone sensor data were observed. Red markers indicate detected change points. Monitoring was restarted following each detected change point.}
\label{fig:realdata_cp}
\end{figure}

Figure \ref{fig:realdata_cp} displays the change points detected by the proposed procedure for each participant. After excluding one participant with insufficient follow-up, the analysis included 43 sensor-data segments from 40 participants. Of these, 31 segments included all 13 mobility features, while the remaining 12 segments included 11 of the 13 mobility features.

VC* identified 86 change points across 34 of the 40 participants included in the analysis. Because monitoring was restarted after each detected change point, participants could contribute multiple change points during follow-up. Change points were identified across participants with varying lengths of follow-up and occurred throughout the monitoring period. 

Overall, these results demonstrate the practical implementation of the proposed calibration procedure in a real-world smartphone monitoring setting.

\section{Discussion}
In this work, we propose a simulation-based procedure for controlling false positives in online change point detection. In many monitoring settings, testing is performed sequentially over time using statistics that are highly dependent due to overlapping data. Standard multiple testing corrections, such as Bonferroni or Šidák, do not account for this dependence structure and can be overly conservative in practice. Our approach directly targets control of a sequential family-wise error rate over a moving window of monitoring days, which aligns more closely with how online monitoring procedures are used in practice.

Simulation results show that the proposed procedure controlled the sFWER at the target level across a range of settings. In contrast, the unadjusted procedure produced inflated false alarm rates, while Bonferroni and Šidák corrections were more conservative than the proposed approach. These findings suggest that accounting for the dependence induced by repeated testing can improve error control without requiring overly stringent thresholds.

Although this work was motivated by mHealth applications, the proposed framework is not specific to smartphone data. Similar multiple testing challenges arise whenever monitoring is performed repeatedly over time and decisions must be made sequentially. As a result, the proposed calibration procedure may also be useful in other settings involving online monitoring.

There are several limitations to this work. First, the proposed procedure relies on simulations, which can be computationally intensive. A large number of simulations may be needed to obtain stable cutoff estimates, particularly when estimating more extreme quantiles of the monitoring statistic. In practice the computational cost in a daily online setting is likely negligible, but for more frequent testing intervals computation could become a concern.

Second, the proposed calibration procedure assumes that the data are stable under the null hypothesis of no change point. This assumption may not hold in practice, particularly in settings where changes may occur gradually over time rather than through a single abrupt shift corresponding to a change point. Future work could investigate the robustness of the proposed calibration procedure to gradual changes in the data over time.


Finally, the mHealth study presented here was intended primarily to illustrate how the proposed procedure may be implemented using passively collected smartphone data. Since the procedure was applied retrospectively and there is no gold standard for identifying true behavioral changes, it is not possible to determine whether detected change points correspond to clinically meaningful events. Future work could evaluate the proposed procedure prospectively in a true online setting and investigate the relationship between detected change points and clinical outcomes.

\section*{Funding}

This work was funded in part by the A.E. Foundation and the Penn-CHOP Lifespan Brain Institute.

\section*{Supplementary Material}
Supplementary material is available in the supplementary appendix accompanying this preprint.

\section*{Conflict of Interest} 
None declared.

\bibliographystyle{plainnat}
\bibliography{references}

\pagebreak

\begin{table}[ht]
\centering
\caption{Estimated rejection rates across simulation settings. When $\tau = 0$, values correspond to Type I error; when $\tau > 0$, values correspond to the empirical probability of detecting at least one change point within the monitoring window.}
\label{tab:rejection_combined}
\resizebox{\textwidth}{!}{
\begin{tabular}{ccc|cc|cc|cc|cc}
\hline
$\alpha_\Delta$ & $\Delta$ & Method 
& \multicolumn{2}{c|}{$\tau=0$}
& \multicolumn{2}{c|}{$\tau=0.5$}
& \multicolumn{2}{c|}{$\tau=1.0$}
& \multicolumn{2}{c}{$\tau=1.5$} \\
& &
& $p=1$ & $p=10$
& $p=1$ & $p=10$
& $p=1$ & $p=10$
& $p=1$ & $p=10$ \\
\hline

\multirow{8}{*}{0.2}
& \multirow{4}{*}{7}
& Proposed     & 0.20 & 0.23 & 0.57 & 0.69 & 0.66 & 0.85 & 0.69 & 0.92 \\
& & Unadjusted   & 0.18 & 0.18 & 0.38 & 0.50 & 0.49 & 0.71 & 0.54 & 0.81 \\
& & Bonferroni   & 0.09 & 0.09 & 0.31 & 0.40 & 0.40 & 0.62 & 0.46 & 0.74 \\
& & Šidák        & 0.10 & 0.10 & 0.31 & 0.43 & 0.41 & 0.64 & 0.47 & 0.76 \\
\cline{2-11}

& \multirow{4}{*}{14}
& Proposed     & 0.22 & 0.20 & 0.48 & 0.56 & 0.57 & 0.76 & 0.63 & 0.84 \\
& & Unadjusted   & 0.31 & 0.26 & 0.55 & 0.63 & 0.64 & 0.81 & 0.68 & 0.89 \\
& & Bonferroni   & 0.11 & 0.10 & 0.35 & 0.39 & 0.45 & 0.61 & 0.52 & 0.72 \\
& & Šidák        & 0.12 & 0.10 & 0.37 & 0.39 & 0.47 & 0.62 & 0.53 & 0.73 \\
\hline

\multirow{8}{*}{0.1}
& \multirow{4}{*}{7}
& Proposed     & 0.08 & 0.10 & 0.43 & 0.55 & 0.52 & 0.74 & 0.58 & 0.83 \\
& & Unadjusted   & 0.18 & 0.18 & 0.38 & 0.50 & 0.49 & 0.71 & 0.54 & 0.81 \\
& & Bonferroni   & 0.04 & 0.05 & 0.22 & 0.29 & 0.33 & 0.52 & 0.38 & 0.65 \\
& & Šidák        & 0.04 & 0.05 & 0.22 & 0.29 & 0.33 & 0.52 & 0.38 & 0.65 \\
\cline{2-11}

& \multirow{4}{*}{14}
& Proposed     & 0.11 & 0.11 & 0.35 & 0.40 & 0.45 & 0.63 & 0.51 & 0.83 \\
& & Unadjusted   & 0.31 & 0.26 & 0.55 & 0.63 & 0.64 & 0.81 & 0.68 & 0.89 \\
& & Bonferroni   & 0.06 & 0.06 & 0.27 & 0.28 & 0.37 & 0.49 & 0.44 & 0.62 \\
& & Šidák        & 0.06 & 0.06 & 0.27 & 0.28 & 0.37 & 0.49 & 0.44 & 0.62 \\
\hline

\end{tabular}
}
\end{table}

\pagebreak

\pagebreak

\end{document}